# Foundation for Frequent Pattern Mining Algorithms' Implementation

Prof. Paresh Tanna[#1], Dr. Yogesh Ghodasara[*2]

[#1]School of Engineering – MCA Department, RK. University, Rajkot, Gujarat, India
[*2]College of Information Tech., Anand Agriculture University, Anand, Gujarat, India

*Abstract*— As with the development of the IT technologies, the amount of accumulated data is also increasing. Thus the role of data mining comes into picture. Association rule mining becomes one of the significant responsibilities of descriptive technique which can be defined as discovering meaningful patterns from large collection of data. The frequent pattern mining algorithms determine the frequent patterns from a database. Mining frequent itemset is very fundamental part of association rule mining. Many algorithms have been proposed from last many decades including majors are Apriori, Direct Hashing and Pruning, FP-Growth, ECLAT etc. The aim of this study is to analyze the existing techniques for mining frequent patterns and evaluate the performance of them by comparing Apriori and DHP algorithms in terms of candidate generation, database and transaction pruning. This creates a foundation to develop newer algorithm for frequent pattern mining.

*Keywords*— Association rule, Frequent pattern mining, Apriori, DHP, Foundation Implementation Study

## I. INTRODUCTION

Automated data collection tools and mature database technology lead to tremendous amounts of data accumulated and/or to be analysed in databases, data warehouses, and other information repositories [7]. We are drowning in data, but starving for knowledge! What is the solution for this problem? I think its Data Mining - Mining interesting knowledge (rules, regularities, patterns, constraints) from data in large databases. Data mining refers to the use of sophisticated data analysis tools to discover previously unknown, valid patterns and relationships in large data sets.[1]"

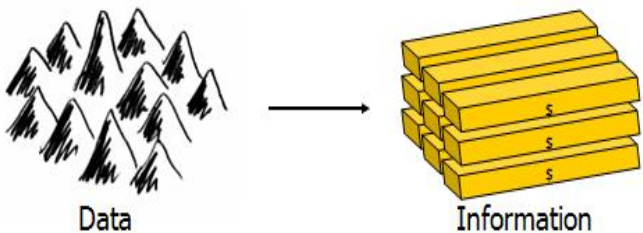

Fig. 1  Data to Information with Data Mining[1]

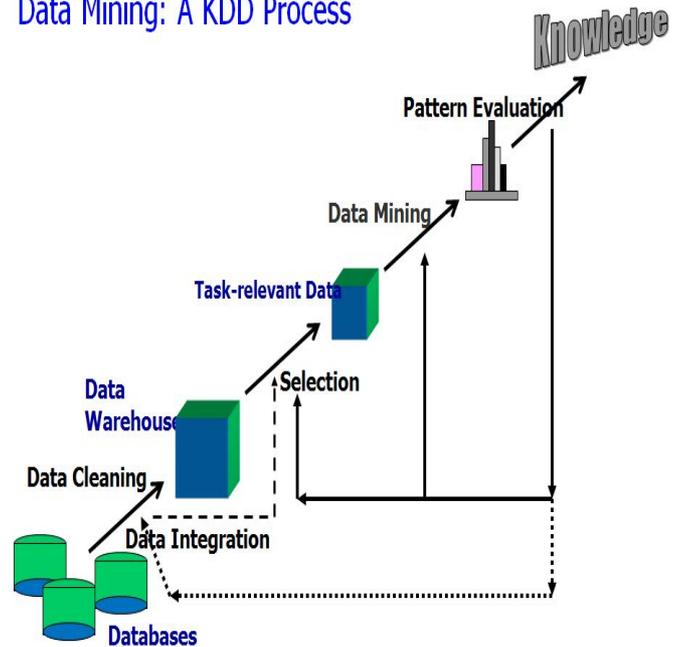

Fig. 2  Data Mining – A KDD process[1]

Frequent Patterns are patterns (such as itemsets) that appear in a data set frequently:  A set of items, like milk and bread, that appear frequently together in a transaction dataset is a frequent itemsets. Frequent pattern mining searches for recurring relationships in a given data set. Researcher can focus on frequent patterns mining like Frequent itemsets from the small and/or from the large amount of data, where the data are either transactional or relational[7]. So many applications are there which we can be considered as frequent pattern applications like Supermarket for product placement & special promotions, Websearch for which keywords often occur together in webpages, Health care for frequent sets of symptoms for a disease, Basically works for all data that can be represented as a set of examples/objects having certain properties like patient / symptoms, movies / ratings, web pages / keywords, basket / products etc. Considering Market Basket Analysis we can find that Market basket analysis might tell a retailer that customers often purchase shampoo and conditioner together, so putting both items on promotion at the same time would not create a significant increase in profit, while a promotion involving just one of the items would likely drive sales of the other.

Association rule learning is a popular and well researched method for discovering interesting relations between variables in large databases[6]. It is intended to identify strong rules discovered in databases using different measures of interestingness. Based on the concept of strong rules[2], introduced association rules for discovering regularities between products in large-scale transaction data recorded by point-of-sale (POS) systems in supermarkets. The volume of data is increasing dramatically as the data generated by day-to-day activities. Therefore, mining association rules from massive amount of data in the database is interested for many industries which can help in many business decision making processes, such as cross-marketing, Basket data analysis, and promotion assortment. The problem of association rule mining is defined as: Let $I = \{i_1, i_2,.....,i_n\}$ be a set of n binary attributes called items. Let $D = = \{t_1, t_2,.....,t_m\}$ be a set of transactions called the database. Each transaction in D has a unique transaction ID and contains a subset of the items in I. A rule is defined as an implication of the form X==>Y where X, Y $\subseteq$ I and X $\cap$ Y = θ. The sets of items (for short itemsets) X and Y are called antecedent (left-hand-side or LHS) and consequent (right-hand-side or RHS) of the rule respectively. To illustrate the concepts, we use a small example from the supermarket domain. The set of items is I = {milk, bread, butter, beer}. An example rule for the supermarket could be {butter, bread} ==> {milk} meaning that if butter and bread are bought, customers also buy milk. In short we summaries association rule as given:  Given database of transactions and each transaction is a list of items(purchased by a customer in a visit) then find all rules that correlate the presence of one set of items with that of another set of item. Example 98% of people who purchase tires and auto accessories also get automotive services done[2].

## II. FREQUENT PATTERN MINING ALGORITHMS

Hundreds of algorithms have been proposed for sparse/dense data, many rows/columns, data fits/does not fit in memory etc. Among these we can filter out most useful methods which we can categorize them as scalable methods for mining frequent patterns. Scalable mining methods: Four major approaches are: Apriori : Fast Algorithms for Mining Association Rules[2], Direct Hashing and Pruning (DHP) : An Effective Hash-Based Algorithm for Mining Association Rules[3], Frequent pattern growth (FP – Growth) : Mining Frequent Patterns without Candidate Generation: A Frequent-Pattern Tree Approach[4], Vertical data format approach (ECLAT): New Algorithms for Fast Discovery of Association Rules[5]

### A. *Apriori: A Candidate Generation-and-Test Approach*

Apriori is a classic algorithm for frequent itemset mining and association rule learning over transactional databases [2]. It proceeds by identifying the frequent individual items in the database and extending them to larger and larger item sets as long as those item sets appear sufficiently often in the database. The frequent item sets determined by Apriori can be used to determine association rules which highlight general trends in the database: this has applications in domains such as market basket analysis.

Apriori is designed to operate on databases containing transactions (for example, collections of items bought by customers, or details of a website frequentation). Each transaction is seen as a set of items (an itemset). Given a threshold C, the Apriori algorithm identifies the itemsets which are subsets of at least C transactions in the database. Apriori uses a "bottom up" approach, where frequent subsets are extended one item at a time (a step known as candidate generation), and groups of candidates are tested against the data. The algorithm terminates when no further successful extensions are found. Apriori uses breadth-first search and a Hash tree structure to count candidate item sets efficiently [2]. It generates candidate item sets of length k from item sets of length k-1. Then it prunes the candidates which have an infrequent sub pattern. According to the downward closure lemma, the candidate set contains all frequent k-length item sets. After that, it scans the transaction database to determine frequent item sets among the candidates. In short, it finds the frequent itemsets : the sets of items that have minimum support and a subset of a frequent itemset must also be a frequent itemset i.e., if {AB} is a frequent itemset, both {A} and {B} should be a frequent itemset. Also iteratively find frequent itemsets with cardinality from 1 to k (k-itemset) with two step process: Join Step: $C_k$ is generated by joining $L_{k-1}$ with itself and Prune Step:  Any (k-1)-itemset that is not frequent cannot be a subset of a frequent k-itemset. Apriori follows the following method: (i) initially, scan DB once to get frequent 1-itemset, (ii) generate length (k+1) candidate itemsets from length k frequent itemsets, (iii) test the candidates against DB and finally (iv) terminate when no frequent or candidate set can be generated. We can note down that any subset of large itemset is large therefore to find large k-itemset: create candidates by combining large k-1 itemsets, delete those that contain any subset that is not large. Example of Generating Candidates Let $L_3$={abc, abd, acd, ace, bcd} the we can have self-joining: $L_3*L_3$ abcd  from abc and abd , acde from acd and ace. Also we can have pruning: Pruning: acde is removed because ade is not in $L_3$ and  $C_4$ will be {abcd}.

*Apriori Algorithm Implementation Summary using java sample code:*

$C_k$: Candidate itemset of size k
$L_k$ : frequent large itemset of size k
find_frequent_itemsets(long m_sup)
```
{
    int k=1;   // Initially k=1
    min_sup = m_sup;
    LinkedList<FrequentItem> L;
    LinkedList<FrequentItem> C=null;
    L = Find_frequent_1_itemsets();
    while(L.size() >= 1)// Line No: 2
```

```
      {
        C = apriori_gen(L,k);  //Generate new k-itemsets candidates
        C=CandidateSupportCount(C,k); //Find the support of all the candidates
          if(C.size()<=0)
            break;
          L=FindLargeItemsets(C); // Take only those with support over minsup
          if(L.size()<=0)
            break;
          k++;
      }
      aps.txtaResult.setText(aps.txtaResult.getText()+outputstr);
  }

//Candidate Generation :
apriori_gen(L,k)
{
    LinkedList<FrequentItem> C = new LinkedList<FrequentItem>();
    for(int i=0;i<L.size();i++)
    {
      for(int j=i+1; j<L.size();j++)
      {
        if(IsPosibleInter(L.get(i).itemset, L.get(j).itemset, k-1))
        {
          int temp = L.get(j).itemset.lastIndexOf(",")+1;
          FrequentItem c = new
             FrequentItem(L.get(i).itemset+","
              + L.get(j).itemset.substring(temp),0);
          if(!Has_Infrequent_Subset(c,L,k))
            C.addLast(c);
        }
      }
    }
    return C;
}
```

Considering an example for joining and pruning : Let $L_3$ = { {1 2 3}, {1 2 4}, {1 3 4}, {1 3 5}, {2 3 4} } After joining : { {1 2 3 4}, {1 3 4 5} } and After pruning : {1 2 3 4} since {1 4 5} and {3 4 5} are not in $L_3$.

Also Apriori algorithm can be modified to improve its efficiency (computational complexity) by hashing, removal of transactions that do not contain frequent itemsets, sampling of the data, partitioning of the data, and mining frequent itemsets without generation of candidate itemsets.

### B. *The DHP Algorithm (Direct Hashing and Pruning) – Improvement approach towards Apriori*

DHP can be used for efficient large itemset generation. It has two major features: efficient generation for large itemsets and effective reduction on transaction database. It uses hashing technique. In particular, for the large 2-itemsets, where the number of candidate large itemsets generated by DHP is, in orders of magnitude, smaller than that of by Apriori method[3]. Thus improving the performance bottleneck of the whole process. It Uses pruning technique to reduce the size of the database progressively[6].

Hashing is used to reduce the size of the candidate k-itemsets, i.e., itemsets that are generated from frequent itemsets from iteration k-1, $C_k$, for k>1. For instance, when scanning D to generate $L_1$ from the candidate 1-temsets in $C_1$, we can at the same time generate all 2-itemsets for each transaction, hash (map) them into different buckets of the hash table structure and increase the corresponding bucket counts[1]. A 2-itemset, which corresponding bucket count is below the support threshold, cannot be frequent and thus we can remove it from the candidate set $C_2$. In this way we reduce the number of candidate 2-itemsets that must be examined to obtain $L_2$. It finds the *frequent itemsets* : the sets of items that have minimum support - a subset of a frequent itemset must also be a frequent itemset  i.e., if {*AB*} is a frequent itemset, both {*A*} and {*B*} should be a frequent itemset.  DHP uses the technique due to this which is more powerful than Apriori i.e. Candidate large 2-itemsets are huge - DHP trims them using hashing and transaction database is huge that one scan per iteration is costly - DHP prunes both number of transactions and number of items in each transaction after each iteration[6].

*Hash Table Construction*
Consider two items sets, all items are numbered as I1, I2, …$I_n$. For any pair (x, y), has according to  Hash function bucket # = h({x y}) = ((order of x)*10+(order of y)) % 7. Example: Items = A, B, C, D, E,  Order  = 1, 2, 3  4, 5,  then H({C, E})= (3*10 + 5)% 7 = 0. Thus, {C, E} belong to bucket 0.

*How to trim candidate itemsets*
In k-iteration, hash all "appearing"  k+1 itemsets in a hashtable, count all the occurrences of an itemset in the correspondent bucket. In k+1 iteration, examine each of the candidate itemset to see if its correspondent bucket value is above the support (necessary condition)[3].

|  |  |  |  |  |  |  |
|--|--|--|--|--|--|--|
|  |  |  |  |  |  |  |

 (I) In trasaction  (A, C, D) , a single candidate AC is found in C2. Occurrence frequencies of all the items are : a[0] = 1, a[1] = 1, a[2] = 0. Since all the values of a[i] are less than k (k=2), this transaction is deemed not useful for generating large 3-itemsets and thus discarded. (II) In transaction(A, B, C, E), has four candidate 2-items (AC, BC, BE, CE) found in C2. Occurrence frequencies of all the items are : a[0] = 1, a[1] = 2, a[2] = 2, a[3] = 2. Since all the values of a[0] are less than k (k=2), and remaining are >=2, this transaction will be reduced to (B, C, E) and A is thus discarded.

*Implementation Evaluation Comparison*

| Apriori Result | | | K | DHP Result | | |
|---|---|---|---|---|---|---|
| Transactions/ Itemsets | Analysis | No. of Rows in Database Table (After Ck) | | Transactions/ Itemsets | Analysis | No. of Rows in Database Table (After Ck) |
| A,C,D<br>B,C,E<br>A,B,C,E<br>B,E | 4 - transactions considered for study | 4 | | A,C,D<br>B,C,E<br>A,B,C,E<br>B,E | 4 - transactions considered for study | 4 |
| ====== C1 ===== | | | | | | |
| A/ 2<br>B/ 3<br>C/ 3<br>D/ 1<br>E/ 3 | 5 - itemsets for C1 | 4 | 1 | A/ 2<br>B/ 3<br>C/ 3<br>D/ 1<br>E/ 3 | 5 - itemsets for C1 | 4 |
| ====== L1 ===== | | | | | | |
| A/ 2<br>B/ 3<br>C/ 3<br>E/ 3 | 4 - itemsets selected for L1 | 4 | | A/ 2<br>B/ 3<br>C/ 3<br>E/ 3 | 4 - itemsets selected for L1 | 4 |
| ====== C2 ===== | | | | | | |
| A,B/ 1<br>A,C/ 2<br>A,E/ 1<br>B,C/ 2<br>B,E/ 3<br>C,E/ 2 | 6 - itemsets for C2 | 4 | | A,C/ 2<br>B,C/ 2<br>B,E/ 3<br>C,E/ 2 | 4 - itemsets for C2 | 2 |
| ====== L2 ===== | | | | | | |
| A,C/ 2<br>B,C/ 2<br>B,E/ 3<br>C,E/ 2 | 4 - itemsets selected for L2 | 4 | | A,C/ 2<br>B,C/ 2<br>B,E/ 3<br>C,E/ 2 | 4 - itemsets selected for L2 | 2 |
| ====== C3 ===== | | | | | | |
| B,C,E/ 2 | 1 - itemset for C3 | 4 | | B,C,E/ 2 | 1 - itemset for C3 | 0 |
| ====== C3 ===== | | | | | | |
| B,C,E/ 2 | 1 - itemset for L3 | 4 | | B,C,E/ 2 | 1 - itemset for L3 | 0 |

*Comparison of Apriori and DHP*

Apriori behaves like - Generate Candidate Set and Perform Count Support from Database. While DHP behaves in a sequence – Generate candidate set, perform count support from the database and make new hash table using database for the next stage.

*Effective Database Pruning*

Apriori - don't prune database but prune $C_k$ by support counting on the original database, while DHP -Its m*ore efficient support counting can be achieved on pruned database*[3]

### III. CONCLUSION

Apriori is best for frequent pattern mining approach for newer algorithm development. But after implementation you can find some challenges like multiple scans of transaction database, huge number of candidates, tedious workload of support counting for candidates and we can improve Apriori with effective hash-based algorithm for the candidate itemset generation i.e. a two phase transaction database pruning and much more efficient ( time & space ) than Apriori algorithm. Following implementation statics we can find that for the below example 6-itemsets for C2 & 4-transactions for database in Apriori while 4-itemsets for C2 & only 2-transactions for database in DHP.

REFERENCES

[1] Data Mining: Concepts and Techniques, Jiawei Han and Micheline Kamber, MORGAN KAUFMANN PUBLISHER, An Imprint of Elsevier

[2] R. Agrawal and S. Srikant, "Fast Algorithms for Mining Association Rules in Large Databases", Proceedings of the 20th International Conference on Very Large Data Bases, September 1994.

[3] J. Park, M. Chen and Philip Yu, "An Effective Hash-Based Algorithm for Mining Association Rules", Proceedings of ACM Special Interest Group of Management of Data, ACM SIGMOD'95, 1995.

[4] Han, Pei & Yin, "Mining Frequent Patterns without Candidate Generation: A Frequent-Pattern Tree Approach", Data Mining and Knowledge Discovery, Volume 8, Issue 1 , pp 53-87,2004

[5] M. Zaki, S. Parthasarathy, M. Ogihara, and W. Li, " New Algorithms for Fast Discovery of Association Rules", Proc. 3rd ACM SIGKDD Int. Conf. on Knowledge Discovery and Data Mining (KDD'97, Newport Beach, CA), 283-296 AAAI Press, Menlo Park, CA, USA 1997

[6] Shruti Aggarwal, Ranveer Kaur, "Comparative Study of Various Improved Versions of Apriori Algorithm", International Journal of Engineering Trends and Technology (IJETT) - Volume4Issue4- April 2013

[7] Agrawal, R., T. Imielin´ski, and A. Swami (1993). Mining association rules between sets of items in large databases. In Proceedings of the 1993 ACM SIGMOD International Conference on Management of Data, SIGMOD '93, New York, NY, USA, pp. 207–216. ACM.